\documentclass[aps,pra,twocolumn,amsmath,amssymb,nofootinbib,superscriptaddress]{revtex4}

\newcommand{\bra}[1]{\langle#1|}
\newcommand{\ket}[1]{|#1\rangle}

\usepackage[pdftex]{graphicx}
\usepackage{mathrsfs}
\usepackage{comment}
\usepackage{epstopdf}%This allowed me to load up eps files
\usepackage{subfig}
\usepackage{setspace}
\usepackage{color}
\usepackage[colorlinks]{hyperref}

\begin{document}

\bibliographystyle{apsrev}

\title{Quantum random walks on congested lattices}

\author{Keith R. Motes}
\email[]{motesk@gmail.com}
\affiliation{Centre for Engineered Quantum Systems, Department of Physics and Astronomy, Macquarie University, Sydney NSW 2113, Australia}

\author{Alexei Gilchrist}
\email[]{alexei@ectropy.info}
\affiliation{Centre for Engineered Quantum Systems, Department of Physics and Astronomy, Macquarie University, Sydney NSW 2113, Australia}

\author{Peter P. Rohde}
\email[]{dr.rohde@gmail.com}
\homepage{http://www.peterrohde.org}
\affiliation{Centre for Engineered Quantum Systems, Department of Physics and Astronomy, Macquarie University, Sydney NSW 2113, Australia}

\date{\today}

\frenchspacing
% ABSTRACT
\begin{abstract}
We consider quantum random walks on congested lattices and contrast them to classical random walks. Congestion is modelled with lattices that contain static defects which reverse the walker's direction. We implement a dephasing process after each step which allows us to smoothly interpolate between classical and quantum random walkers as well as study the effect of dephasing on the quantum walk. Our key results show that a quantum walker escapes a finite boundary dramatically faster than a classical walker and that this advantage remains in the presence of heavily congested lattices. Also, we observe that a quantum walker is extremely sensitive to our model of dephasing.

%The propagation of the walker decreases monotonically with the spatial dephasing rate. We find that the propagation of a walker is \emph{highly} sensitive to small process dephasing rates and quickly localises, causing the quantum random walk to have the characteristics of a classical random walk. Also, spatial dephasing does not lead to localisation like process dephasing does. In other words spatial dephasing preserves the quadratic spreading behaviour of the walker while process dephasing does not.
\end{abstract}
\maketitle

%INTRODUCTION
\section{Introduction}

Quantum information processing \cite{bib:NielsenChuang00} promises many interesting technologies that are not available today. Perhaps most interesting is the promise for quantum computation, whereby quantum algorithms can be implemented that outperform their classical counterparts. The best known example is Shor's factoring algorithm \cite{bib:Shor97}, which can factor numbers exponentially faster than the best known classical factoring algorithm. Other examples include Grover's database search algorithm \cite{bib:Grover96} and various graph theoretic algorithms \cite{bib:Ambainis2003, PhysRevA.81.052313, PhysRevA.83.042317}. While the technologies to implement these algorithms are not currently available, it is important to study potential routes towards implementing technologies that can implement these algorithms. 

One route to implementing quantum information processing tasks is via quantum random walks \cite{bib:ADZ, bib:AAKV, bib:Kempe08, bib:Salvador12} whereby a particle, such as a photon, `hops' between the vertices in a lattice. In this paper the effects of a congested, or obstructed, lattice on a quantum random walk (QRW) are studied and compared to a classical random walk (CRW). The quantum walkers also suffer a dephasing process as they propagate. This study provides insight into how random errors in the lattice and dephasing affect the dynamics of random walks and the robustness of certain quantum features. In our model, congestion refers to where the lattice through which the walker  propagates has defects. These random defects are like blocked streets that the walker encounters and has to back out of on the next step. These defects are stationary during the evolution of the random walk, though we average over many such random lattices. Dephasing occurs when the state decoheres and is implemented via a dephasing channel acting after each step. In the limit of full dephasing the quantum walk becomes a classical walk, so that dephasing also allows us to interpolate between the classical and quantum regimes. For an experimental implementation of dephasing see Broome \emph{et al.} \cite{bib:PhysRevLett.104.153602}, and for related theoretical work on quantum walks with phase damping see Lockhart \emph{et al.} \cite{bib:lockhart2013performance}.

For characterising the resulting probability distributions for the quantum and classical random walks we use the variance and the `escape probability', that is the probability that the walker escapes a finite region of the lattice, or more picturesquely, the probability that the walker `beats the traffic'.

%QUANTUM RANDOM WALKS
\section{Quantum Random Walks}
A QRW describes the evolution of a quantum particle through a given topological structure represented as a $d$ dimensional lattice. In a classical random walk, the walker probabilistically follows edges through a lattice to step to an adjacent vertex. In a QRW on the other hand, the walker spreads as a superposition of different paths through the graph. Physically, the walker can be a wide range of quantum particles, though of particular interest is the photon as photons are readily produced, manipulated and measured using off-the-shelf components in the laboratory. Photons have found widespread use in quantum information processing, most notably linear optics quantum computing (LOQC) \cite{bib:KLM01}. These technologies provide the topological structure for implementing a QRW. They also allow for multi-photon QRWs \cite{bib:increasing12}, which increases the dimensionality of the walk. For a further review on QRWs see Refs. \cite{bib:ADZ, bib:AAKV, bib:Kempe08, bib:Salvador12}, and see Refs. \cite{bib:Hagai08, bib:Schreiber10, bib:Broome10, bib:Peruzzo10, bib:Schreiber11b, bib:Matthews11, bib:Owens11, bib:Schreiber12, Sansoni12} for the numerous optical demonstrations of elementary QRWs that have been performed.
  
\subsection{Quantum random walk formalism}
To illustrate our QRW formalism we present the details for a one-dimensional discrete QRW on an unbounded lattice without any defects. The state of a one-dimensional QRW at any given time has the form,
\begin{equation} \label{eq:State}
\ket{\Psi}=\sum_{x, c} \gamma_{x,c} \ket{x, c},
\end{equation}
where $x \in [-t_\mathrm{max},t_\mathrm{max}]$ represents the position of the particle; $t_\mathrm{max}$ represents the total number of time steps and thus the size of the lattice; $c \in \{-1,1\}$ is the coin value that tells the walker whether to evolve to the left ($c=-1$) or right ($c=1$); and $|\gamma_{x,c}|^{2}$ is the probability amplitude at a given position and coin value. Since there are two coin values for each position, the probably that the walker is at position $x$ is given by,
\begin{equation} \label{eq:probX}
P(x)=|\gamma_{x,-1}|^{2}+|\gamma_{x,1}|^{2}.
\end{equation}

The one-dimensional walker begins at some specified input state $\ket{\Psi(0)}=\ket{x_{0},c_{0}}$ before it begins to evolve at time $t=0$, where $x_{0}$ and $c_{0}$ are the starting position and starting coin value respectively. Typically $x_{0}$ is chosen to be the origin. The state then evolves for a finite number of time steps. The evolution is described by two operators: the coin $\hat{C}$ and step $\hat{S}$ operators, 
\begin{eqnarray} 
\hat{C} \ket{x, \pm 1}&=&(\ket{x,1}\pm \ket{x,-1})/\sqrt{2} \label{eq:coin} \\ \nonumber
\hat{S}\ket{x, c}&=&\ket{x+c,c} \label{eq:step}.
\end{eqnarray}
The coin operator takes a state and maps it to a superposition of new states using the Hadamard coin,
\begin{equation} \label{eq:H}
H= \frac{1}{\sqrt{2}} \begin{pmatrix}
1 & 1 \\
1 & -1\\
\end{pmatrix},\\
\end{equation}
exploiting both possible degrees of freedom in the coin while maintaining the same position. Next, the step operator $\hat{S}$ moves the walker to an adjacent position according to the value of $c$. $\hat{C}$ and $\hat{S}$ act on the state at every time step and thus the full evolution of the system is given by,
\begin{equation}
\ket{\Psi(t)} = (\hat{S}\cdot \hat{C})^{t}\ket{\Psi(0)}.
\end{equation}
If the walker begins at the origin or on an even lattice position then, as the walker evolves, it lies on odd positions for odd time steps and on even positions for even time steps. Thus, as the walker evolves, the allowed locations for the walker oscillate between even and odd sites.

It is straightforward to generalise Eq.~\ref{eq:State} to multiple dimensions by expanding the Hilbert space. For example, a two-dimensional walk would have the form
\begin{equation}
\ket{\Psi^{(2)}}=\sum_{x,y,c_{x},c_{y}} \gamma_{x,y,c_{x},c_{y}} \ket{x,y, c_{x},c_{y}},
\end{equation}
where $x\in [-t_\mathrm{max},t_\mathrm{max}]$ and $y\in [-t_\mathrm{max},t_\mathrm{max}]$ denote the two spatial dimensions, $c_{x} \in \{-1,1\}$ indicates for the walker to move left or right, $c_{y} \in \{-1,1\}$ indicates for the walker to move down or up, and the superscript represents the dimension. The coin and step operator can be generalised by taking a tensor product for each respective dimension, or alternately a coin could be employed which entangles the two dimensions. In the case of a spatially separable two-dimensional coin one obtains $\hat{C}^{(2)}=\hat{C}_{x}\otimes \hat{C}_{y}$ and $\hat{S}^{(2)}=\hat{S}_{x}\otimes \hat{S}_{y}$. Likewise, the hadamard coin for two dimensions becomes $H\otimes H$.

After the system evolves, a measurement is made on either the position or the coin degree of freedom yielding the  output probability distribution. With this probability distribution various metrics can be defined to characterise the evolution of the system, which we define next.

%%------------
%\subsection{Generalising our QRW formalism to implement a CRW}
%%------------
%To make a comparison between classical and quantum random walks, we first generalised our QRW formalism to implement a CRW. To do this we modified the QRW coin operator of Eq.~\ref{eq:coin} by using the following left $L$ or right $R$ coins,
%\begin{eqnarray}
%L&=& \begin{pmatrix}
%0 & 1 \\
%0 & 1\\
%\end{pmatrix}\\ \nonumber
%R&=& \begin{pmatrix}
%1 & 0 \\
%1 & 0\\
%\end{pmatrix}.\\ \nonumber
%\end{eqnarray}
%For the one-dimensional case this probabilistically maps the state to either left or right with probability $1/2$, rather than as a superposition. In the two-dimensional case the walker propagates either up, down, left or right with probability $1/4$. Since the classical walker never enters a super-position, there is never more than one term in the state $\ket\Psi$ and the probability distribution is given by a single peak. Thus, to obtain proper statistics, the walk is averaged over many iterations. Once the statistics of the distribution are obtained over many iterations the variance $\sigma^{2}$ and the escape probability $P_\mathrm{esc}$ can be obtained, as described in the next section.

%------------
\subsection{Random Walk Metrics}
%------------
The two common metrics that we use to quantify a QRW are the variance $\sigma^{2}$ and the escape probability $P_\mathrm{esc}$.

\subsubsection{Variance}
%______
The variance $\sigma^{2}$ is a measure of how much the walker has spread out during its evolution. It is defined as,
\begin{equation}
\sigma^2=\sum_{i=1}^{n}{p_{i}(i-\mu)^{2}},
\end{equation}
where $n=2\,t_\mathrm{max}+1$ and $\mu=\sum_{i=1}^{n}{p_{i} i}$. Fig.~\ref{fig:PescVsTime} (top) illustrates the variance versus time for both a QRW and a CRW on a two-dimensional square lattice of size $t_\mathrm{max}=20$. The QRW demonstrates a quadratic rate of spreading across the lattice while the CRW demonstrates a linear rate of spreading. This quadratic spreading is one of the distinguishing features of a QRW compared to the CRW. It forms the basis of some quantum walk algorithms such as the quantum walk search algorithm, which is quadratically faster than the corresponding classical algorithm.

%______
\subsubsection{Escape Probability}
%______
%define the boundary as a square within the walkers' lattice. More formally, if the square lattice is defined by some $t_\mathrm{max}$ then the boundary is the square of width $e=2 \times t_{b}$ where $0\leq t_{b} \leq t_\mathrm{max}$ is how far the boundary is from the origin. 

 The escape probability $P_\mathrm{esc}$ is a measure of how much the walker leaks outside of a certain region on the walker's lattice. To answer this question a boundary must first be defined which depends on the size of the lattice. For the square two-dimensional lattice we let the walker begin at $(x=-t_\mathrm{max},y=0)$ and let the boundary be at $x=t_b$, where $t_b$ is how far the boundary is from the left edge of the lattice. To calculate the escape probability on this square lattice we use,
\begin{equation} \label{eq:Pesc}
P_\mathrm{esc}=\sum_{x\in{x_\mathrm{out}}} \sum_{y}P^{(2)}(x,y),
\end{equation}
where $x_\mathrm{out}$ are the positions outside of the boundary and $P^{(2)}(x,y)$ is the two-dimensional version of Eq.~\ref{eq:probX}.
%**********Figure*****************
\begin{figure}[!htb]
\includegraphics[scale=0.88]{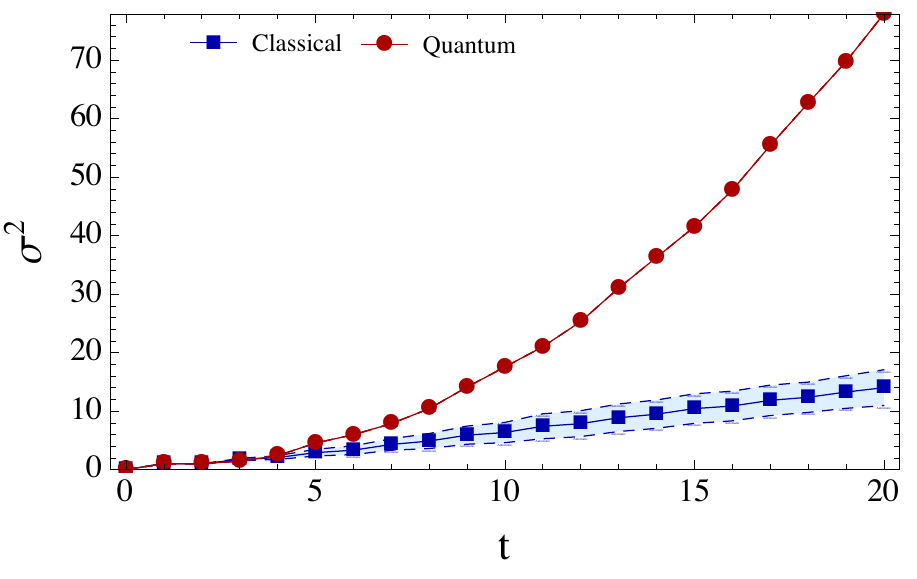}\\
\includegraphics[scale=0.88]{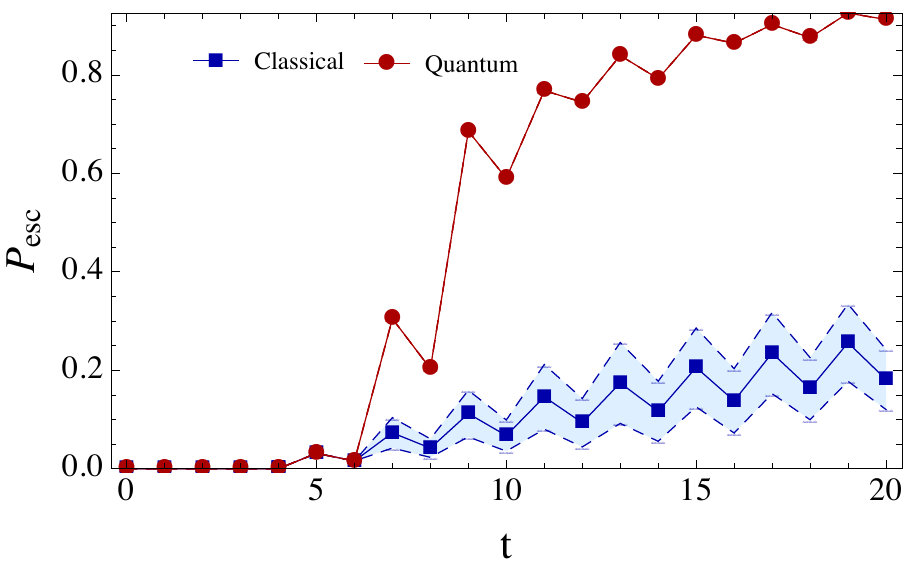}
\caption{\label{fig:PescVsTime} (Top) The variance $\sigma^{2}$ versus time $t$ for the classical and quantum random walk on a two-dimensional square lattice defined by $t_\mathrm{max}=20$. The rate of spreading is quadratic for the quantum case and linear for the classical case. (Bottom) The escape probability $P_\mathrm{esc}$ against time for the classical and quantum random walk on a two-dimensional square lattice defined by $t_\mathrm{max}=20$ with a boundary defined by $t_{b}=4$. In the quantum case, the probability of escape is much larger for any given time after escaping than in the classical case.}
\end{figure}
%***********************************

Fig.~\ref{fig:PescVsTime} (bottom) illustrates $P_\mathrm{esc}$ versus $t$ for both a QRW and a CRW on a square lattice of size $t_\mathrm{max}=20$ with a boundary given by $t_{b}=4$. Here the quantum case exhibits a dramatic jump in escape probability compared to the classical case. This is due to both the faster rate of spreading of the QRW, and to the QRW having larger amplitudes at the tails of its distribution. This dramatic jump is a key feature pointed out in this work that demonstrates an advantage that QRWs have over CRWs. In our study the walker is allowed to walk back into the unescaped region which takes away from the probability that the walker has escaped. This, in conjunction with the fact that the walker occupies alternating even and odd positions as the walker evolves, explains the oscillatory nature of the escape probability.

The two metrics, $\sigma^{2}$ and $P_{\mathrm{esc}}$, are closely related. If the walker has a large spread in its distribution then the walker also has a better chance to fall outside of the escape boundary. They also capture different aspects of the distribution. At any given time step $t$ during the evolution we can determine the probability distribution with Eq.~\ref{eq:probX} and then calculate these various metrics to be used for quantifying a random walk. Next, we demonstrate how to add spatial defects, which cause congestion, into the walkers' lattice and explore how the variance and escape probability are affected by this lattice congestion. 

% SPATIAL DEPHASING
\section{Lattice Congestion}

Lattice congestion is a model of defects in a medium. For the QRW and CRW the medium is the walkers' lattice and the defects are modelled as blocked pathways where the walker has to enter the pathway to realise it is blocked and then reverse out on the next step. This model is closely related to percolation theory which models defects as missing lattice nodes. For a detailed introduction on percolation theory see \cite{bib:Shante1971, bib:blanc1986introduction}. It is generally modelled on a $d$ dimensional lattice with a given geometry such as a square, triangle or honeycomb. Regardless of geometry, the lattice consists of two components: $\textit{sites}$ and $\textit{bonds}$. A site is a point on the lattice and a bond is the connection between the sites. These components give two strategies for introducing the random fluctuations that define percolation theory: \emph{site percolations} and \emph{bond percolations}, where the term `percolations' refers to the defects on the lattice. In site percolation the lattice points exist with probability $p\in [0,1]$. When a point does not exist it is a defect in the lattice. In bond percolation the positions in a lattice are fixed while the bonds between the positions exist with probability $p$. The model in this paper is a variant of site percolation whereby the walker can occupy any site, but with probability $1-p$ will find an obstruction and reverse direction upon hitting the respective site. We expect the same percolation characteristics such as percolation thresholds to exists in the underlying lattice that the walkers are exploring. For a two-dimensional square lattice with site percolations that most closely resemble the lattice used in this paper, the percolation threshold is $p_c\approx 0.6$ \cite{bib:PhysRevB.40.636}. Values of $p$ higher than this threshold produce long-range connectedness in the lattice. 

To generate a lattice with spatial defects a matrix of coin operators is constructed. The matrix is the same size as the lattice and each position in the matrix corresponds to a spatial position on the lattice. The coin operator corresponding to a given position then determines the behaviour of the walker. The coin operators are defined as either a Hadamard coin (Eq.~\ref{eq:H}), if the site is present, or a bit-flip coin $X$ if the site contains a defect,
\begin{equation} \label{eq:bitflip}
X= \begin{pmatrix}
0 & 1\\
1 & 0\\
\end{pmatrix}.
\end{equation}
For the two dimensional case $X\otimes X$ is used. As the quantum or classical walker evolves it will walk into these defects that signify congested points of the lattice. Upon reaching a defect the walker reverses direction, thus slowing the walker's rate of spread. In this manuscript we define $p$ as the probability that the site is not a defect; therefore, the probability that a site is a defect is $1-p$.  

\subsection{CRW on a congested lattice}

The lattices we are considering contain randomly distributed defects, or points of congestion that impede the walkers progress. Questions such as what is the probability that there is an open path from one side of the lattice to the other, are answered by \emph{percolation theory}. There are many known applications for percolation theory \cite{bib:sahimi1994applications}. A common example is asking whether a liquid can flow through a porous material. If enough pores (or sites) exist then the liquid can make it through. Another example is whether or not an electric current can flow through some medium where conductive sites are spread throughout some insulator. If enough conductive sites are present then a path will exist through the medium. For a more detailed account of percolation theory see \cite{bib:grimmett1999percolation, bib:Shante}.

Within the congested lattice we examine the spread of random walkers. Defects have the effect of reducing the rate of spread of the walker, or stopping it entirely if the lattice is so congested that there is no escape possible from the region the walker finds itself in. 

%************
\subsection{QRW on a congested lattice}

%Most importantly, Viv Kendon \textit{et. al.} shows how a one-dimensional quantum walk on a percolated lattice with a barrier spreads with quantum tunnelling introduced. They found here that this leads to decoherence and a $1/\sqrt{t}$ scaling in the spread time. For two-dimensions they show how the spreading rate \dots Nonetheless, we explore percolation theory in a different context with no tunnelling probability. 

\begin{figure}[]
\includegraphics[scale=0.88]{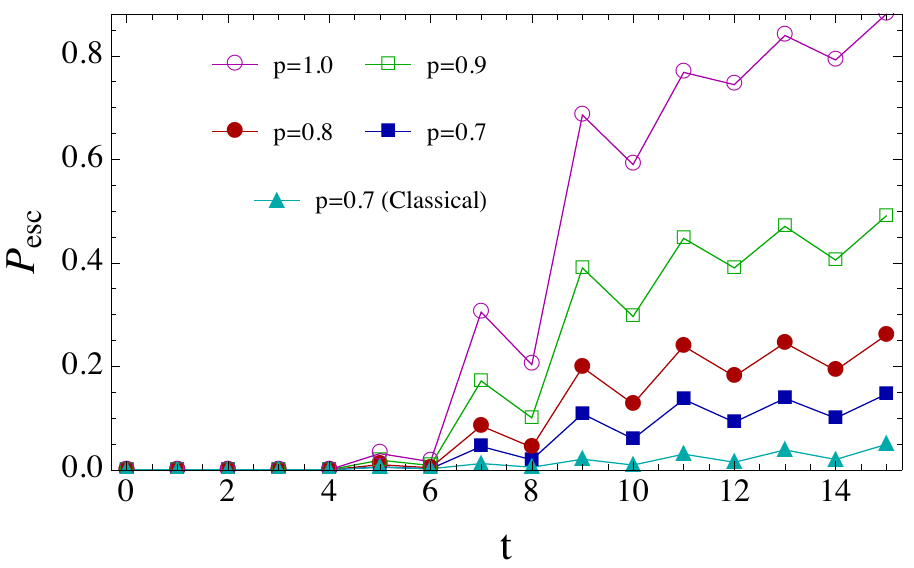}
\caption{\label{fig:PescVsTVaryingP} The escape probability $P_{\mathrm{esc}}$ plotted as a function of time $t$ for varying congestion probabilities $1-p$ on a two-dimensional square lattice of size $t_\mathrm{max}=15$ with a boundary of $t_{b}=4$ and input state $\ket{\Psi}=\ket{-t_\mathrm{max},0,1,1}$. As $p$ decreases the jump in $P_{\mathrm{esc}}$ becomes less prominent. The CRW is presented for $p=0.7$ to illustrate the reduced rate of escape compared to a QRW with the same value of congestion.}
\end{figure}

Classically, the state can only move in one direction at a time while quantum mechanically the state spreads in a superposition of every direction simultaneously. As with a classical walker, the quantum walker escapes the bounded region more often if there are less defects. The significance of the quantum walker is both the quadratic behaviour which means that it escapes more rapidly than the classical walker, and that the resulting probability distribution has more weight in the tails. For a review of work done on QRW with percolation see \cite{bib:PhysRevLett.108.230505, bib:VivKendon2010}. Fig.~\ref{fig:PescVsTVaryingP} shows the escape probability $P_{\mathrm{esc}}$ versus time $t$ for varying values of congestion probability $1-p$ on a lattice of size $t_\mathrm{max}=15$ with an input state of $\ket{\Psi}=\ket{-t_\mathrm{max},0,1,1}$ and boundary $t_b=4$. For $p=1$ there is no congestion present and the $P_{\mathrm{esc}}$ metric experiences a sudden jump from $t=4$ to $t=5$. This is because the QRW has most of its amplitude in its tails as it evolves. When $p$ decreases and the lattice becomes more and more congested the sudden jump is still present at the same value of $t$ but with a much smaller amplitude. Also, for a congestion of $p=0.7$ we present both the QRW and the CRW. This shows that QRWs retain their advantage over CRWs in the presence of 
heavy congestion. Note that the percolation threshold is around $p\approx 0.6$, below which we expect that on average there is no clear route across the graph. 

%**********Figure*****************
%***********************************

%Remember that $p_{c}$ is defined as the probability where above which points A and B are almost certainly connected while below which the points A and B are almost certainly not connected. 

%Thus a different metric is needed to quantify the system. Some possible metrics are the variance of the walker spreading with time or the probability for a walker to escape a specified region. 

%Calculating this for every position yields a probability distribution that we take the variance $\sigma^2$ to quantify our QRW. The variance shows how the QRW is advantageous over the classical random walk (CRW). The variance evolves against time with ballistic spreading $\sigma^2 = O(t^{2})$ while CRW's exhibit linear spreading $\sigma^2 = O(t)$. This means that a QRW can implement algorithms significantly faster. Although QRW's and CRW's have many similarities they ask fundamentally different questions when percolation is introduced. In fact, there is little literature on QRW's walking on percolated lattices \cite{bib:VivKendon2010}.

 %DEPHASING
\section{Dephasing}

Next, we consider what happens to a QRW subject to dephasing. Dephasing represents decoherence caused by the environment which can be related to measurement errors caused by thermal fluctuations, white noise, photons interfering with the quantum walker, etc. To explore this we first introduce a model of dephasing and characterise it with our two metrics: variance and escape probability. 

Consider a quantum walk where after each step, each state in the basis has probability $p_{\mathrm{d}}$ of acquiring a $\pi$ phase flip. We can model this process as choosing to apply one of a set $\{F_j\}$ of unitary matrices covering all the combinations of $\pm 1$ on the diagonal. If $F_j$ has $s$ -1's on the diagonal we choose it with probability $p_{\mathrm{d}}^s(1-p_{\mathrm{d}})^{m-s}$. 

The probability of a particular sequence will be the product of the probabilities of the $F_j$ appearing in the sequence since they are independently chosen at each step. If $\rho_\mathrm{seq}$ is the final pure density matrix appearing with probability $p_\mathrm{seq}$, then in general the final state of the system is described by
\begin{equation}
    \rho = \sum_\mathrm{seq} p_\mathrm{seq} \rho_\mathrm{seq}.
\end{equation}
That is, for any POVM element $E$ we have 
\begin{equation}
    \sum_\mathrm{seq} p_\mathrm{seq} \mathrm{Tr}\{E \rho_\mathrm{seq}\} = \mathrm{Tr}\{E \rho\}.
\end{equation}

We algorithmically implement dephasing by randomly flipping the signs of individual kets in the walker's superposition state with probability $p_{\mathrm{d}}$, and average the results of any measurement at the end of a large number of runs. This in effect samples from the  distribution represented by $\rho$ and is automatically weighted by the probability of a given sequence. 

That this whole process represents dephasing is not immediately obvious. To see it, we first rewrite $\rho$ as the vector $\ket{\rho}$ using the \emph{vec} operation which simply stacks its columns on top of each other. Using the identity $\ket{ABC}=C^T\!\!\otimes\!A\,\ket{B}$ for any three square matrices $A$, $B$, and $C$; then grouping the terms that turn up, we can write 
\begin{equation} \label{eq:vecrho}
    \ket{\rho} = \ldots \sum_k p_k D_k U\sum_j p_j D_j U\sum_i p_i D_i U\ket{\rho_0}
\end{equation}
where $D_j = F_j^*\!\otimes\!F_j=F_j^{\otimes 2}$, $U$ represents the step and coin operations, and $\ket{\rho_0}$ is the vectorised initial density matrix. This shows that after each step we apply the process described by the dynamical matrix
\begin{equation}
    D = \sum_j p_j F_j^{\otimes 2}.
\end{equation}

The matrices $F_j$ are diagonal so we write the diagonal as a vector denoted by $\ket{f}_j$, so that the diagonal of $F_j^{\otimes 2}$ is $\ket{f}_j\ket{f}_j$. Since $\ket{f}_j$ has only real entries we can rearrange it into the matrix $\ket{f}_j\bra{f}$. We can do a similar arrangement with $D$ so that,
\begin{equation}
    \ket{d}\!\bra{d} =  \sum_j p_j \ket{f}_j\bra{f}.
\end{equation}
It's worthwhile pausing and noting what this matrix represents. From Eq.~\ref{eq:vecrho} we can see that the diagonal of $D$ multiplies the elements of the vectorised $\ket{\rho}$. Hence when we arrange the values into a matrix, the entries of $\ket{d}\!\bra{d}$ multiply the corresponding entries in $\rho$.

The first thing to note is that this matrix is symmetric. We will denote the entries of $\ket{f}_j$ by $f_k$ and drop the reference $j$ for clarity. The diagonals of $\ket{f}_j\bra{f}$ are of the form $f_k^2=1$ and since $\sum_j p_j = 1$ the diagonal of $\ket{d}\bra{d}$ is unity and the process does not change the amplitudes of the states. The off-diagonals are of the form $f_rf_s$ where $r\ne s$ and their sum over $j$ has the value
\begin{equation}
    (1-p_d)^2+p_d^2-2(1-p_d)p_d = (1-2p_d)^2.
\end{equation}
The terms on the left are the probabilities that both $f_r$ and $f_s$ are positive, both negative, or one of each respectively.
Each of these terms is multiplied by the binomial sum of the probabilities of all the combinations of $\pm 1$ on all the other elements of $\ket{f}_j$ and not $r$ or $s$, which evaluates to 1. Note that this result holds for any dimension. In summary, the map that is performed by $D$ multiplies every off-diagonal element of $\rho$ by $(1-2p_{\mathrm{d}})^2$. This is a dephasing map.

If $p_{\mathrm{d}}=0$ none of the signs are flipped, and if $p_{\mathrm{d}}=1$ all of the signs are flipped. Since the QRW is invariant under a global phase flip, these two extremes reproduce an ideal QRW. When $0<p_{\mathrm{d}} <1$ dephasing is introduced into the system. A value of $p_{\mathrm{d}}=1/2$ corresponds to complete dephasing which causes the walker to behave classically. The classical results in this paper were produced by using our QRW code with a value of $p_{\mathrm{d}}=1/2$. This was checked with purely classical code to verify that we are indeed obtaining a CRW. 

If we imagine an inefficient measurement of the quantum walk at every step where it is projectively measured with probability $p_{\mathrm{m}}$ or otherwise left alone, this map would describe dephasing by a dynamical matrix which multiplies all the off diagonal elements of $\rho$ by $1-p_{\mathrm{m}}$. So our dephasing process is equivalent to a measurement performed with a probability $p_{\mathrm{m}}=4(1-p_{\mathrm{d}})p_{\mathrm{d}}$.

%**********Figure*****************
\begin{figure}[]
\includegraphics[scale=0.5]{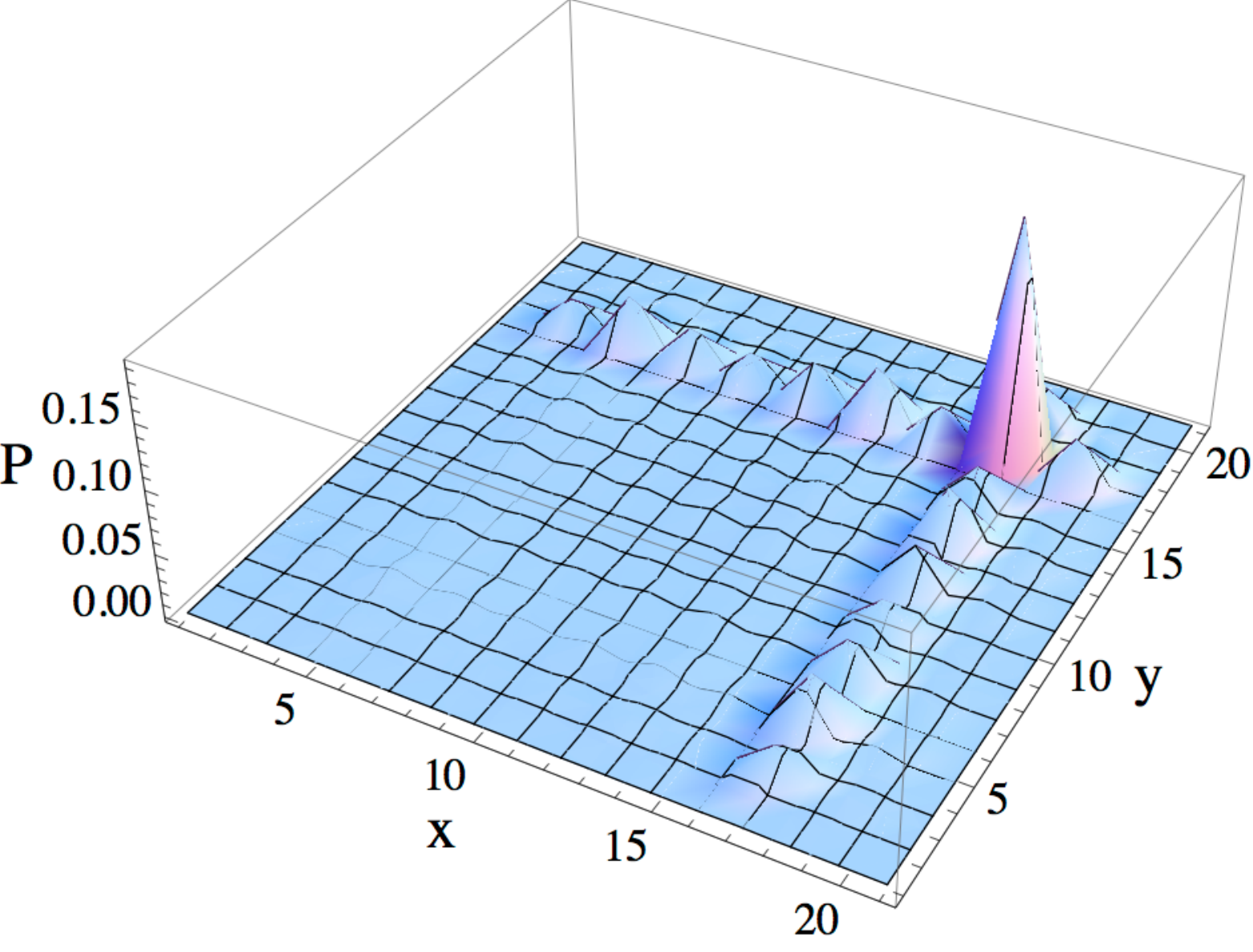}\\
\includegraphics[scale=0.5]{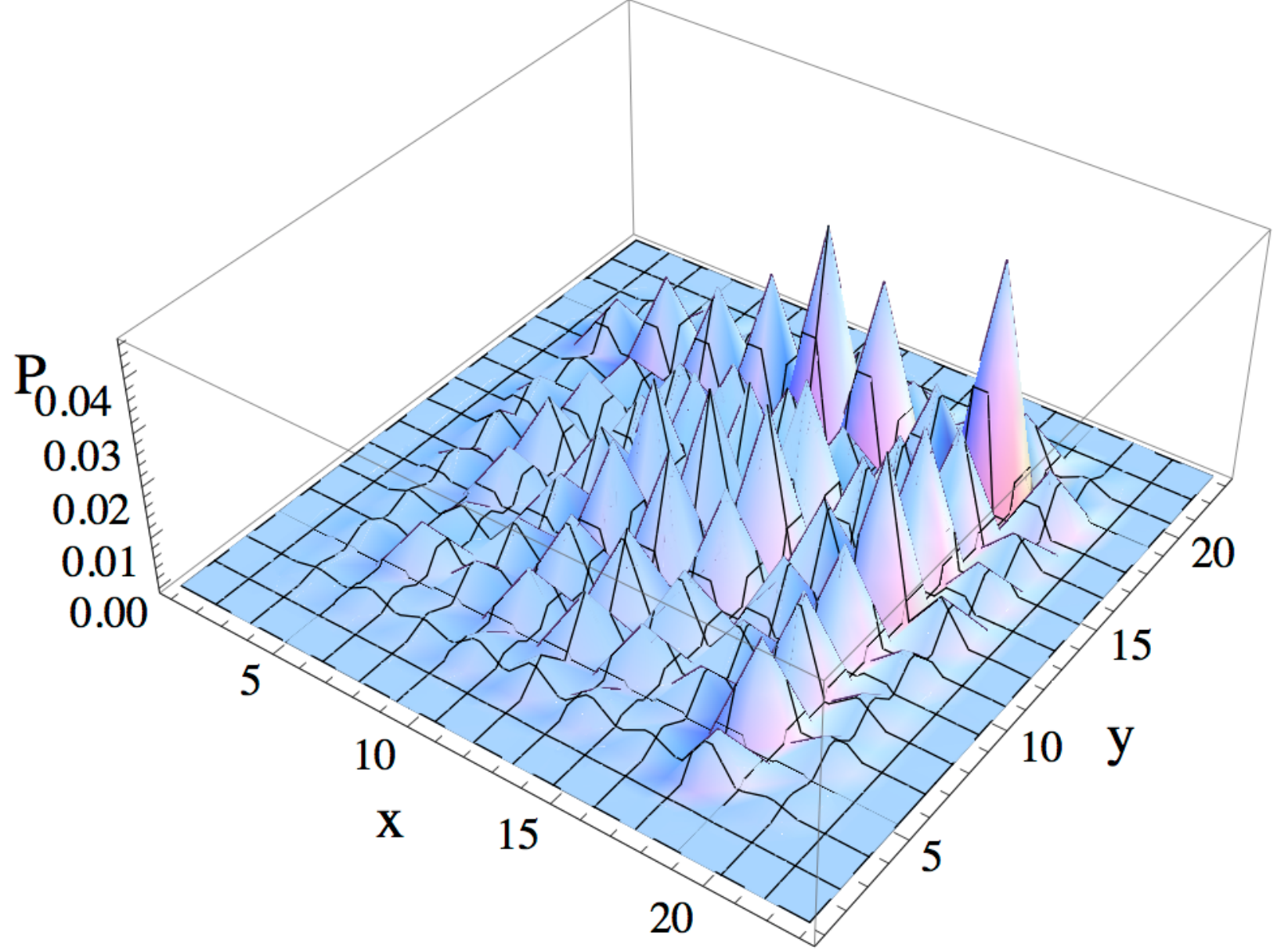}
\caption{\label{fig:AndersonNoDephasing} The QRW probability distribution shown at the final time step over a two-dimensional square lattice defined by $t_\mathrm{max}=10$ using a non-symmetrical input state of $\ket{\Psi}=\ket{0,0,1,1}$. (Top) The QRW with no defects or dephasing always yields this deterministic probability distribution. (Bottom) The same QRW is averaged over many iterations with a small dephasing probability of $p_{\mathrm{d}}=0.00015$. It has a similar probability distribution but is approaching classical statistics.}
\end{figure}
%***********************************

To illustrate the effect of dephasing in our model we first plot the probability distribution at the final time step of the QRW with no dephasing as shown in Fig.~\ref{fig:AndersonNoDephasing} (Top). Here we employ an asymmetric input state of $\ket{\Psi}=\ket{0,0,1,1}$ and let the number of time steps be $t_\mathrm{max}=10$. This distribution has one main peak near the edge of the lattice in the same direction that the walker was initialised in and is completely deterministic. This is in contrast to what occurs when dephasing is introduced. Fig.~\ref{fig:AndersonNoDephasing} (Bottom) shows the same probability distribution again but with a dephasing probability of $p_{\mathrm{d}}=0.00015$. With this small value of dephasing the distribution retains most of its quantum behaviour as in Fig.~\ref{fig:AndersonNoDephasing} (Top) but it begins to approach the statistics of a classical distribution.

\begin{figure}[]
\includegraphics[scale=0.5]{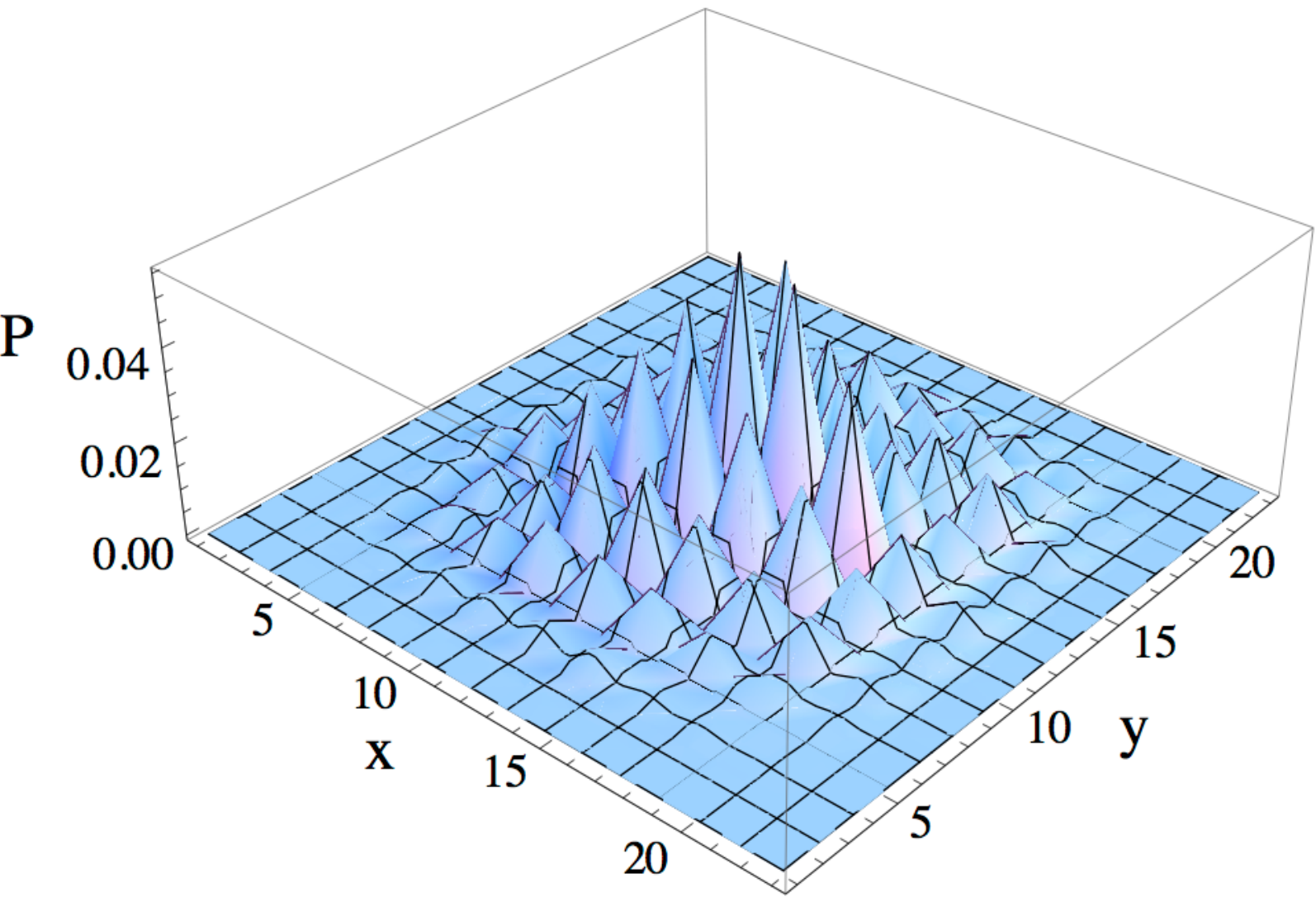}
\caption{\label{fig:ThreeDBinomial} The QRW probability distribution shown at the final time step over a two-dimensional square lattice defined by $t_\mathrm{max}=10$ using a non-symmetrical input state of $\ket{\Psi}=\ket{0,0,1,1}$ and a dephasing probability of $p_{\mathrm{d}}=0.0005$. For $p_{\mathrm{d}}\gtrsim 0.0005$ the probability distribution becomes centred around the origin which corresponds to the probability distribution of a classical random walk.}
\end{figure}

In this work, dephasing is a method for introducing quantum decoherence to the quantum random walk. With sufficiently strong dephasing the quantum walk becomes identical to a classical random walk. We find that the quantum walk is extremely sensitive to this model of dephasing. When $p_{\mathrm{d}}\gtrsim 0.0005$ the probability distribution becomes strongly centered around the origin, which corresponds to the probability distribution of a CRW. This is shown in Fig.~\ref{fig:ThreeDBinomial} for the same input state and lattice size as in Fig.~\ref{fig:AndersonNoDephasing} but with $p_{\mathrm{d}}=0.0005$. 
%**********Figure*****************
%***********************************
This means that just a few sign errors during a walk can cause the whole QRW to behave classically and lose some of its quantum advantages.  By incrementing the dephasing through this interval we can smoothly interpolate between quantum and classical random walks, which is a key feature of this work. 

The extreme sensitivity of our dephasing model is surprising as it is far more sensitive than the dephasing observed by Lockhart \emph{et al.} \cite{bib:lockhart2013performance}; however, there are several notable differences that would account for having different sensitivities. Firstly, Lockhart \emph{et al.} apply phase damping on only the coin degree of freedom whereas we apply it to both the position \emph{and} coin degrees of freedom. Secondly the parametrisation of the dephasing is significantly different, in our approach it corresponds to an inefficient measurement model, where which a certain probability $p_{\mathrm{m}}=4(1-p_{\mathrm{d}})p_{\mathrm{d}}$ the quantum walker is projectively measured in both position and coin. 

%Spatial \& Process Dephasing Combined
\section{Congestion \& Dephasing Combined}
\begin{figure}[]
\includegraphics[scale=0.5]{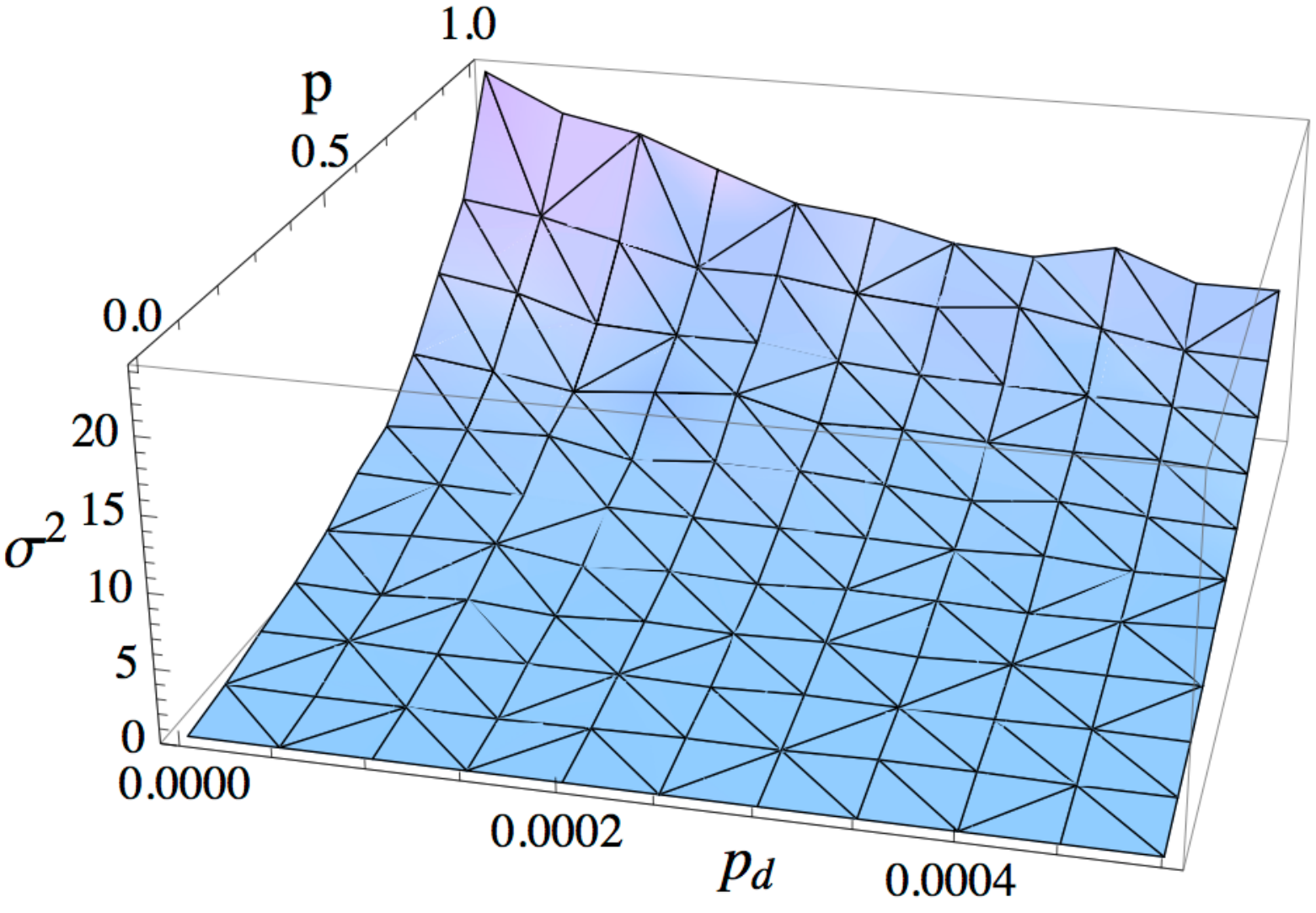}
\caption{\label{fig:DepPercVar} The variance obtained at the final time step plotted against the dephasing probability $p_{\mathrm{d}}$ and the congestion probability $1-p$ for a quantum random walk on a square two-dimensional lattice of size given by $t_\mathrm{max}=10$ and an input state of $\ket{\Psi}=\ket{0,0,1,1}$. The propagation of the walker decreases monotonically with the congestion rate.}
\end{figure}

Next we combine congestion and dephasing and examine the joint effects. Fig.~\ref{fig:DepPercVar} shows the variance obtained at the final time step of the QRW as a function of the dephasing probability $p_{\mathrm{d}}$ and the defect probability $p$ on a two-dimensional square lattice of size given by $t_\mathrm{max}=10$ and an input state of $\ket{\Psi}=\ket{0,0,1,1}$. A monotonic decrease is observed in the variance for a given $p$ as $p_{\mathrm{d}}$ is increased. Also, for any given congestion probability the variance of a QRW decreases as the dephasing probability increases.

\begin{figure}[]
\includegraphics[scale=0.88]{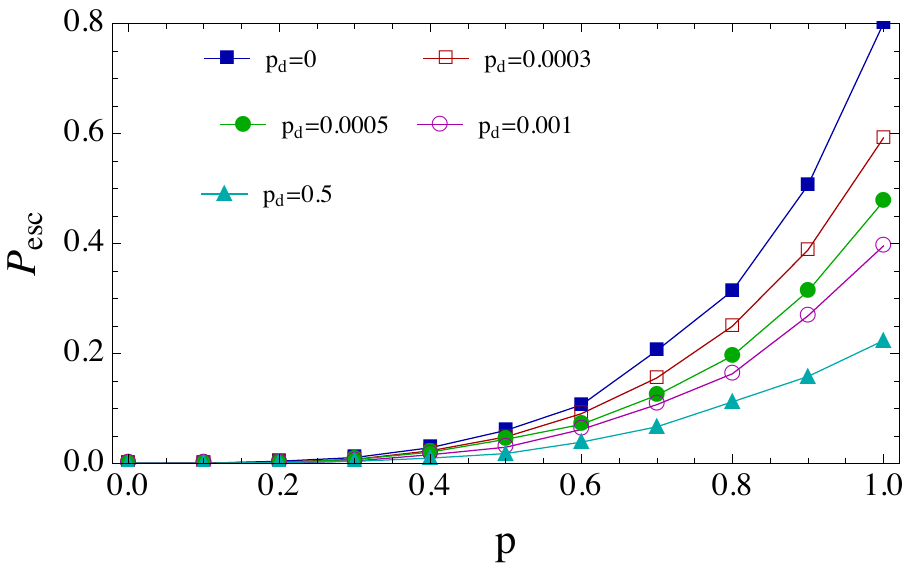}
\caption{\label{fig:PescPPd} The escape probability $P_{\mathrm{esc}}$ with boundary $t_{b}=2$ plotted as a function of congestion probability $1-p$ for varying values of dephasing probabilities $p_{\mathrm{d}}$ on a two-dimensional square lattice of size given by $t_\mathrm{max}=10$ with an input state of $\ket{\Psi}=\ket{-t_\mathrm{max},0,1,1}$. With low dephasing the quantum walker has a larger chance to escape the boundary. As $p_{\mathrm{d}}$ increases the QRW enters the classical regime and the escape probability becomes linear.}
\end{figure}

Fig.~\ref{fig:PescPPd} shows $P_\mathrm{esc}$ with boundary $t_{b}=2$ as a function of congestion probability $1-p$ for varying values of dephasing probabilities $p_{\mathrm{d}}$ on a two-dimensional square lattice defined by $t_\mathrm{max}=10$ with input state $\ket{\Psi}=\ket{-t_\mathrm{max},0,1,1}$. When $p_{\mathrm{d}}=0$ the walk is fully quantum so more of the probability distribution escapes the boundary. When dephasing is increased process errors are introduced, reducing $P_{\mathrm{esc}}$ for any given value of $p$. With dephasing values of $p_{\mathrm{d}}\gtrsim 0.0005$ the QRW enters the classical regime. This suggests that small dephasing rates are large enough to inhibit the quantum advantages of a QRW. Note that (in this case) for $p\lesssim0.3$ none of the probability amplitude escapes the boundary for any value of $p_{\mathrm{d}}$ simply because there are too many defects in the graph.

%**********Figure*****************
%***********************************

%**********Figure*****************
%***********************************

%CONCLUSION
\section{Conclusion}

Quantum random walks are a promising route towards quantum information processing, exhbiting many unique features compared to the classical random walk. In the classical context, walks on percolated lattices (i.e. lattices containing congestion) have been well studied. We have considered the analogous situation in the quantum context. We defined a mapping between quantum and classical walks, via the coin operator, to allow for a direct comparison of the two. Then we introduced a model for adding static defects to the underlying lattice via the introduction of bit-flip coins. These defects inhibit the spread of the classical and quantum walker, reducing the escape probability and variance metrics. Most interestingly, we found that as a quantum random walk evolves it will suddenly and dramatically escape a finite boundary. It maintains this property even in the presence of congestion.

We also introduce a dephasing error model. Dephasing errors are errors caused by the environment on the quantum walker as it evolves. In the limit of large dephasing the quantum random walk spatially localises and behaves like a classical random walk. The spread of the walker is very sensitive to small amounts of dephasing in our dephasing model.
 
We also studied the effects of spatial defects and dephasing together on the propagation of the walker. We found that a monotonic decrease is observed in the variance for any given congestion probability as the dephasing probability is increased. Our results indicate that a quantum walker on a lattice with defects still exhibits a quadratic rate of spread. Thus, as the quadratic spread of quantum walks is one of the key features that make them applicable to quantum information processing applications, such as the quantum search algorithm, quantum walks on congested lattices remain interesting. 

%Importantly, unlike classical percolation theory, where phase-transitions in the path existence metric are observed, the percolated quantum walk does not appear to exhibit a phase-transition when measured by the quantum metrics (variance and escape probability). This might be suggestive of genuinely different behaviour between the quantum and classical cases, or alternately it might be an artefact of the finite lattice sizes we consider, as percolation thresholds are an asymototic result obtained in the limit of large lattice sizes. 

\singlespacing
% ACKNOWLEDGMENTS
\begin{acknowledgments}
This research was conducted by the Australian Research Council Centre of Excellence for Engineered Quantum Systems (Project number CE110001013). We thank Matthew Broome for helpful discussions.
\end{acknowledgments}

% BIBLIOGRAPHY
\bibliography{paper}

\begin{thebibliography}{31}
\expandafter\ifx\csname natexlab\endcsname\relax\def\natexlab#1{#1}\fi
\expandafter\ifx\csname bibnamefont\endcsname\relax
  \def\bibnamefont#1{#1}\fi
\expandafter\ifx\csname bibfnamefont\endcsname\relax
  \def\bibfnamefont#1{#1}\fi
\expandafter\ifx\csname citenamefont\endcsname\relax
  \def\citenamefont#1{#1}\fi
\expandafter\ifx\csname url\endcsname\relax
  \def\url#1{\texttt{#1}}\fi
\expandafter\ifx\csname urlprefix\endcsname\relax\def\urlprefix{URL }\fi
\providecommand{\bibinfo}[2]{#2}
\providecommand{\eprint}[2][]{\url{#2}}

\bibitem[{\citenamefont{Nielsen and Chuang}(2000)}]{bib:NielsenChuang00}
\bibinfo{author}{\bibfnamefont{M.~A.} \bibnamefont{Nielsen}} \bibnamefont{and}
  \bibinfo{author}{\bibfnamefont{I.~L.} \bibnamefont{Chuang}},
  \emph{\bibinfo{title}{Quantum Computation and Quantum Information}}
  (\bibinfo{publisher}{Cambridge University Press, Cambridge},
  \bibinfo{year}{2000}).

\bibitem[{\citenamefont{Shor}(1997)}]{bib:Shor97}
\bibinfo{author}{\bibfnamefont{P.~W.} \bibnamefont{Shor}},
  \bibinfo{journal}{SIAM J. Comput.} \textbf{\bibinfo{volume}{26}},
  \bibinfo{pages}{1484} (\bibinfo{year}{1997}).

\bibitem[{\citenamefont{Grover}(1996)}]{bib:Grover96}
\bibinfo{author}{\bibfnamefont{L.~K.} \bibnamefont{Grover}},
  \bibinfo{journal}{Proc. 28th Annual ACM Symp. on the Theory of Computing} p.
  \bibinfo{pages}{212} (\bibinfo{year}{1996}).

\bibitem[{\citenamefont{AMBAINIS}(2003)}]{bib:Ambainis2003}
\bibinfo{author}{\bibfnamefont{A.}~\bibnamefont{AMBAINIS}},
  \bibinfo{journal}{International Journal of Quantum Information}
  \textbf{\bibinfo{volume}{01}}, \bibinfo{pages}{507} (\bibinfo{year}{2003}).

\bibitem[{\citenamefont{Gamble et~al.}(2010)\citenamefont{Gamble, Friesen,
  Zhou, Joynt, and Coppersmith}}]{PhysRevA.81.052313}
\bibinfo{author}{\bibfnamefont{J.~K.} \bibnamefont{Gamble}},
  \bibinfo{author}{\bibfnamefont{M.}~\bibnamefont{Friesen}},
  \bibinfo{author}{\bibfnamefont{D.}~\bibnamefont{Zhou}},
  \bibinfo{author}{\bibfnamefont{R.}~\bibnamefont{Joynt}}, \bibnamefont{and}
  \bibinfo{author}{\bibfnamefont{S.~N.} \bibnamefont{Coppersmith}},
  \bibinfo{journal}{Phys. Rev. A} \textbf{\bibinfo{volume}{81}},
  \bibinfo{pages}{052313} (\bibinfo{year}{2010}).

\bibitem[{\citenamefont{Berry and Wang}(2011)}]{PhysRevA.83.042317}
\bibinfo{author}{\bibfnamefont{S.~D.} \bibnamefont{Berry}} \bibnamefont{and}
  \bibinfo{author}{\bibfnamefont{J.~B.} \bibnamefont{Wang}},
  \bibinfo{journal}{Phys. Rev. A} \textbf{\bibinfo{volume}{83}},
  \bibinfo{pages}{042317} (\bibinfo{year}{2011}).

\bibitem[{\citenamefont{Aharonov et~al.}(1993)\citenamefont{Aharonov,
  Davidovich, and Zagury}}]{bib:ADZ}
\bibinfo{author}{\bibfnamefont{Y.}~\bibnamefont{Aharonov}},
  \bibinfo{author}{\bibfnamefont{L.}~\bibnamefont{Davidovich}},
  \bibnamefont{and} \bibinfo{author}{\bibfnamefont{N.}~\bibnamefont{Zagury}},
  \bibinfo{journal}{Phys. Rev. A} \textbf{\bibinfo{volume}{48}},
  \bibinfo{pages}{1687} (\bibinfo{year}{1993}).

\bibitem[{\citenamefont{Aharonov et~al.}(2001)\citenamefont{Aharonov, Ambainis,
  Kempe, and Vazirani}}]{bib:AAKV}
\bibinfo{author}{\bibfnamefont{D.}~\bibnamefont{Aharonov}},
  \bibinfo{author}{\bibfnamefont{A.}~\bibnamefont{Ambainis}},
  \bibinfo{author}{\bibfnamefont{J.}~\bibnamefont{Kempe}}, \bibnamefont{and}
  \bibinfo{author}{\bibfnamefont{U.}~\bibnamefont{Vazirani}},
  \bibinfo{journal}{STOC '01 Proceedings of the 33rd ACM symposium on Theory of
  computing} \textbf{\bibinfo{volume}{50}} (\bibinfo{year}{2001}).

\bibitem[{\citenamefont{Kempe}(2003)}]{bib:Kempe08}
\bibinfo{author}{\bibfnamefont{J.}~\bibnamefont{Kempe}},
  \bibinfo{journal}{Cont. Phys.} \textbf{\bibinfo{volume}{44}},
  \bibinfo{pages}{307} (\bibinfo{year}{2003}).

\bibitem[{\citenamefont{Venegas-Andraca}(2012)}]{bib:Salvador12}
\bibinfo{author}{\bibfnamefont{S.~E.} \bibnamefont{Venegas-Andraca}},
  \bibinfo{journal}{QIP} \textbf{\bibinfo{volume}{5}}, \bibinfo{pages}{1015}
  (\bibinfo{year}{2012}).

\bibitem[{\citenamefont{Broome et~al.}(2010{\natexlab{a}})\citenamefont{Broome,
  Fedrizzi, Lanyon, Kassal, Aspuru-Guzik, and
  White}}]{bib:PhysRevLett.104.153602}
\bibinfo{author}{\bibfnamefont{M.~A.} \bibnamefont{Broome}},
  \bibinfo{author}{\bibfnamefont{A.}~\bibnamefont{Fedrizzi}},
  \bibinfo{author}{\bibfnamefont{B.~P.} \bibnamefont{Lanyon}},
  \bibinfo{author}{\bibfnamefont{I.}~\bibnamefont{Kassal}},
  \bibinfo{author}{\bibfnamefont{A.}~\bibnamefont{Aspuru-Guzik}},
  \bibnamefont{and} \bibinfo{author}{\bibfnamefont{A.~G.} \bibnamefont{White}},
  \bibinfo{journal}{Phys. Rev. Lett.} \textbf{\bibinfo{volume}{104}},
  \bibinfo{pages}{153602} (\bibinfo{year}{2010}{\natexlab{a}}).

\bibitem[{\citenamefont{Lockhart et~al.}(2013)\citenamefont{Lockhart,
  Di~Franco, and Paternostro}}]{bib:lockhart2013performance}
\bibinfo{author}{\bibfnamefont{J.}~\bibnamefont{Lockhart}},
  \bibinfo{author}{\bibfnamefont{C.}~\bibnamefont{Di~Franco}},
  \bibnamefont{and}
  \bibinfo{author}{\bibfnamefont{M.}~\bibnamefont{Paternostro}},
  \bibinfo{journal}{arXiv preprint arXiv:1303.5319}  (\bibinfo{year}{2013}).

\bibitem[{\citenamefont{Knill et~al.}(2001)\citenamefont{Knill, Laflamme, and
  Milburn}}]{bib:KLM01}
\bibinfo{author}{\bibfnamefont{E.}~\bibnamefont{Knill}},
  \bibinfo{author}{\bibfnamefont{R.}~\bibnamefont{Laflamme}}, \bibnamefont{and}
  \bibinfo{author}{\bibfnamefont{G.}~\bibnamefont{Milburn}},
  \bibinfo{journal}{Nature (London)} \textbf{\bibinfo{volume}{409}},
  \bibinfo{pages}{46} (\bibinfo{year}{2001}).

\bibitem[{\citenamefont{Rohde et~al.}(2013)\citenamefont{Rohde, Schreiber,
  Stefanak, Jex, Gilchrist, and Silberhorn}}]{bib:increasing12}
\bibinfo{author}{\bibfnamefont{P.~P.} \bibnamefont{Rohde}},
  \bibinfo{author}{\bibfnamefont{A.}~\bibnamefont{Schreiber}},
  \bibinfo{author}{\bibfnamefont{M.}~\bibnamefont{Stefanak}},
  \bibinfo{author}{\bibfnamefont{I.}~\bibnamefont{Jex}},
  \bibinfo{author}{\bibfnamefont{A.}~\bibnamefont{Gilchrist}},
  \bibnamefont{and}
  \bibinfo{author}{\bibfnamefont{C.}~\bibnamefont{Silberhorn}},
  \bibinfo{journal}{J. Comp. and Th. Nanosc. (in press)}
  (\bibinfo{year}{2013}).

\bibitem[{\citenamefont{Perets et~al.}(2008)\citenamefont{Perets, Lahini,
  Pozzi, Sorel, Morandotti, and Silberberg}}]{bib:Hagai08}
\bibinfo{author}{\bibfnamefont{H.~B.} \bibnamefont{Perets}},
  \bibinfo{author}{\bibfnamefont{Y.}~\bibnamefont{Lahini}},
  \bibinfo{author}{\bibfnamefont{F.}~\bibnamefont{Pozzi}},
  \bibinfo{author}{\bibfnamefont{M.}~\bibnamefont{Sorel}},
  \bibinfo{author}{\bibfnamefont{R.}~\bibnamefont{Morandotti}},
  \bibnamefont{and}
  \bibinfo{author}{\bibfnamefont{Y.}~\bibnamefont{Silberberg}},
  \bibinfo{journal}{Phys. Rev. Lett.} \textbf{\bibinfo{volume}{100}},
  \bibinfo{pages}{170506} (\bibinfo{year}{2008}).

\bibitem[{\citenamefont{Schreiber et~al.}(2010)\citenamefont{Schreiber,
  Cassemiro, Poto{\u c}ek, G{\' a}bris, Mosley, Andersson, Jex, and
  Silberhorn}}]{bib:Schreiber10}
\bibinfo{author}{\bibfnamefont{A.}~\bibnamefont{Schreiber}},
  \bibinfo{author}{\bibfnamefont{K.~N.} \bibnamefont{Cassemiro}},
  \bibinfo{author}{\bibfnamefont{V.}~\bibnamefont{Poto{\u c}ek}},
  \bibinfo{author}{\bibfnamefont{A.}~\bibnamefont{G{\' a}bris}},
  \bibinfo{author}{\bibfnamefont{P.~J.} \bibnamefont{Mosley}},
  \bibinfo{author}{\bibfnamefont{E.}~\bibnamefont{Andersson}},
  \bibinfo{author}{\bibfnamefont{I.}~\bibnamefont{Jex}}, \bibnamefont{and}
  \bibinfo{author}{\bibfnamefont{C.}~\bibnamefont{Silberhorn}},
  \bibinfo{journal}{Phys. Rev. Lett.} \textbf{\bibinfo{volume}{104}},
  \bibinfo{pages}{050502} (\bibinfo{year}{2010}).

\bibitem[{\citenamefont{Broome et~al.}(2010{\natexlab{b}})\citenamefont{Broome,
  Fedrizzi, Lanyon, Kassal, Aspuru-Guzik, and White}}]{bib:Broome10}
\bibinfo{author}{\bibfnamefont{M.~A.} \bibnamefont{Broome}},
  \bibinfo{author}{\bibfnamefont{A.}~\bibnamefont{Fedrizzi}},
  \bibinfo{author}{\bibfnamefont{B.~P.} \bibnamefont{Lanyon}},
  \bibinfo{author}{\bibfnamefont{I.}~\bibnamefont{Kassal}},
  \bibinfo{author}{\bibfnamefont{A.}~\bibnamefont{Aspuru-Guzik}},
  \bibnamefont{and} \bibinfo{author}{\bibfnamefont{A.~G.} \bibnamefont{White}},
  \bibinfo{journal}{Phys. Rev. Lett.} \textbf{\bibinfo{volume}{104}},
  \bibinfo{pages}{153602} (\bibinfo{year}{2010}{\natexlab{b}}).

\bibitem[{\citenamefont{Peruzzo et~al.}(2010)\citenamefont{Peruzzo, Lobino,
  Matthews, Matsuda, Politi, Poulios, Zhou, Lahini, Ismail, W{\" o}rhoff
  et~al.}}]{bib:Peruzzo10}
\bibinfo{author}{\bibfnamefont{A.}~\bibnamefont{Peruzzo}},
  \bibinfo{author}{\bibfnamefont{M.}~\bibnamefont{Lobino}},
  \bibinfo{author}{\bibfnamefont{J.~C.~F.} \bibnamefont{Matthews}},
  \bibinfo{author}{\bibfnamefont{N.}~\bibnamefont{Matsuda}},
  \bibinfo{author}{\bibfnamefont{A.}~\bibnamefont{Politi}},
  \bibinfo{author}{\bibfnamefont{K.}~\bibnamefont{Poulios}},
  \bibinfo{author}{\bibfnamefont{X.-Q.} \bibnamefont{Zhou}},
  \bibinfo{author}{\bibfnamefont{Y.}~\bibnamefont{Lahini}},
  \bibinfo{author}{\bibfnamefont{N.}~\bibnamefont{Ismail}},
  \bibinfo{author}{\bibfnamefont{K.}~\bibnamefont{W{\" o}rhoff}},
  \bibnamefont{et~al.}, \bibinfo{journal}{Science}
  \textbf{\bibinfo{volume}{329}}, \bibinfo{pages}{1500} (\bibinfo{year}{2010}).

\bibitem[{\citenamefont{Schreiber et~al.}(2011)\citenamefont{Schreiber,
  Cassemiro, Potocek, Gabris, Jex, and Silberhorn}}]{bib:Schreiber11b}
\bibinfo{author}{\bibfnamefont{A.}~\bibnamefont{Schreiber}},
  \bibinfo{author}{\bibfnamefont{K.~N.} \bibnamefont{Cassemiro}},
  \bibinfo{author}{\bibfnamefont{V.}~\bibnamefont{Potocek}},
  \bibinfo{author}{\bibfnamefont{A.}~\bibnamefont{Gabris}},
  \bibinfo{author}{\bibfnamefont{I.}~\bibnamefont{Jex}}, \bibnamefont{and}
  \bibinfo{author}{\bibfnamefont{C.}~\bibnamefont{Silberhorn}},
  \bibinfo{journal}{Phys. Rev. Lett.} \textbf{\bibinfo{volume}{106}},
  \bibinfo{pages}{180403} (\bibinfo{year}{2011}).

\bibitem[{\citenamefont{Matthews et~al.}(2011)\citenamefont{Matthews, Poulios,
  Meinecke, Politi, Peruzzo, Ismail, W{\" o}rhoff, Thompson, and
  O'Brien}}]{bib:Matthews11}
\bibinfo{author}{\bibfnamefont{J.~C.~F.} \bibnamefont{Matthews}},
  \bibinfo{author}{\bibfnamefont{K.}~\bibnamefont{Poulios}},
  \bibinfo{author}{\bibfnamefont{J.~D.~A.} \bibnamefont{Meinecke}},
  \bibinfo{author}{\bibfnamefont{A.}~\bibnamefont{Politi}},
  \bibinfo{author}{\bibfnamefont{A.}~\bibnamefont{Peruzzo}},
  \bibinfo{author}{\bibfnamefont{N.}~\bibnamefont{Ismail}},
  \bibinfo{author}{\bibfnamefont{K.}~\bibnamefont{W{\" o}rhoff}},
  \bibinfo{author}{\bibfnamefont{M.~G.} \bibnamefont{Thompson}},
  \bibnamefont{and} \bibinfo{author}{\bibfnamefont{J.~L.}
  \bibnamefont{O'Brien}} (\bibinfo{year}{2011}), \eprint{arXiv:1106.1166}.

\bibitem[{\citenamefont{Owens et~al.}(2011)\citenamefont{Owens, Broome,
  Biggerstaff, Goggin, Fedrizzi, Linjordet, Ams, Marshall, Twamley, Withford
  et~al.}}]{bib:Owens11}
\bibinfo{author}{\bibfnamefont{J.~O.} \bibnamefont{Owens}},
  \bibinfo{author}{\bibfnamefont{M.~A.} \bibnamefont{Broome}},
  \bibinfo{author}{\bibfnamefont{D.~N.} \bibnamefont{Biggerstaff}},
  \bibinfo{author}{\bibfnamefont{M.~E.} \bibnamefont{Goggin}},
  \bibinfo{author}{\bibfnamefont{A.}~\bibnamefont{Fedrizzi}},
  \bibinfo{author}{\bibfnamefont{T.}~\bibnamefont{Linjordet}},
  \bibinfo{author}{\bibfnamefont{M.}~\bibnamefont{Ams}},
  \bibinfo{author}{\bibfnamefont{G.~D.} \bibnamefont{Marshall}},
  \bibinfo{author}{\bibfnamefont{J.}~\bibnamefont{Twamley}},
  \bibinfo{author}{\bibfnamefont{M.~J.} \bibnamefont{Withford}},
  \bibnamefont{et~al.}, \bibinfo{journal}{New. J. Phys.}
  \textbf{\bibinfo{volume}{13}}, \bibinfo{pages}{075003}
  (\bibinfo{year}{2011}).

\bibitem[{\citenamefont{Schreiber et~al.}(2012)\citenamefont{Schreiber,
  G{\'a}bris, Rohde, Laiho, {\v S}tefa{\v n}{\' a}k, Poto{\u c}ek, Jex, and
  Silberhorn}}]{bib:Schreiber12}
\bibinfo{author}{\bibfnamefont{A.}~\bibnamefont{Schreiber}},
  \bibinfo{author}{\bibfnamefont{A.}~\bibnamefont{G{\'a}bris}},
  \bibinfo{author}{\bibfnamefont{P.~P.} \bibnamefont{Rohde}},
  \bibinfo{author}{\bibfnamefont{K.}~\bibnamefont{Laiho}},
  \bibinfo{author}{\bibfnamefont{M.}~\bibnamefont{{\v S}tefa{\v n}{\' a}k}},
  \bibinfo{author}{\bibfnamefont{V.}~\bibnamefont{Poto{\u c}ek}},
  \bibinfo{author}{\bibfnamefont{I.}~\bibnamefont{Jex}}, \bibnamefont{and}
  \bibinfo{author}{\bibfnamefont{C.}~\bibnamefont{Silberhorn}},
  \bibinfo{journal}{Science} \textbf{\bibinfo{volume}{336}},
  \bibinfo{pages}{55} (\bibinfo{year}{2012}).

\bibitem[{\citenamefont{Sansoni et~al.}(2012)\citenamefont{Sansoni, Sciarrino,
  Vallone, Mataloni, Crespi, Ramponi, and Osellame}}]{Sansoni12}
\bibinfo{author}{\bibfnamefont{L.}~\bibnamefont{Sansoni}},
  \bibinfo{author}{\bibfnamefont{F.}~\bibnamefont{Sciarrino}},
  \bibinfo{author}{\bibfnamefont{G.}~\bibnamefont{Vallone}},
  \bibinfo{author}{\bibfnamefont{P.}~\bibnamefont{Mataloni}},
  \bibinfo{author}{\bibfnamefont{A.}~\bibnamefont{Crespi}},
  \bibinfo{author}{\bibfnamefont{R.}~\bibnamefont{Ramponi}}, \bibnamefont{and}
  \bibinfo{author}{\bibfnamefont{R.}~\bibnamefont{Osellame}},
  \bibinfo{journal}{Phys. Rev. Lett.} \textbf{\bibinfo{volume}{108}},
  \bibinfo{pages}{010502} (\bibinfo{year}{2012}).

\bibitem[{\citenamefont{Shante and
  Kirkpatrick}(1971{\natexlab{a}})}]{bib:Shante1971}
\bibinfo{author}{\bibfnamefont{V.~K.} \bibnamefont{Shante}} \bibnamefont{and}
  \bibinfo{author}{\bibfnamefont{S.}~\bibnamefont{Kirkpatrick}},
  \bibinfo{journal}{Advances in Physics} \textbf{\bibinfo{volume}{20}},
  \bibinfo{pages}{325} (\bibinfo{year}{1971}{\natexlab{a}}).

\bibitem[{\citenamefont{Blanc}(1986)}]{bib:blanc1986introduction}
\bibinfo{author}{\bibfnamefont{R.}~\bibnamefont{Blanc}}, in
  \emph{\bibinfo{booktitle}{Contribution of Clusters Physics to Materials
  Science and Technology}} (\bibinfo{publisher}{Springer},
  \bibinfo{year}{1986}), pp. \bibinfo{pages}{425--478}.

\bibitem[{\citenamefont{Yonezawa et~al.}(1989)\citenamefont{Yonezawa, Sakamoto,
  and Hori}}]{bib:PhysRevB.40.636}
\bibinfo{author}{\bibfnamefont{F.}~\bibnamefont{Yonezawa}},
  \bibinfo{author}{\bibfnamefont{S.}~\bibnamefont{Sakamoto}}, \bibnamefont{and}
  \bibinfo{author}{\bibfnamefont{M.}~\bibnamefont{Hori}},
  \bibinfo{journal}{Phys. Rev. B} \textbf{\bibinfo{volume}{40}},
  \bibinfo{pages}{636} (\bibinfo{year}{1989}).

\bibitem[{\citenamefont{Sahimi}(1994)}]{bib:sahimi1994applications}
\bibinfo{author}{\bibfnamefont{M.}~\bibnamefont{Sahimi}},
  \emph{\bibinfo{title}{Applications of percolation theory}}
  (\bibinfo{publisher}{CRC PressI Llc}, \bibinfo{year}{1994}).

\bibitem[{\citenamefont{Grimmett}(1999)}]{bib:grimmett1999percolation}
\bibinfo{author}{\bibfnamefont{G.~R.} \bibnamefont{Grimmett}},
  \emph{\bibinfo{title}{Percolation}}, vol. \bibinfo{volume}{321}
  (\bibinfo{publisher}{Springer}, \bibinfo{year}{1999}).

\bibitem[{\citenamefont{Shante and
  Kirkpatrick}(1971{\natexlab{b}})}]{bib:Shante}
\bibinfo{author}{\bibfnamefont{V.~K.} \bibnamefont{Shante}} \bibnamefont{and}
  \bibinfo{author}{\bibfnamefont{S.}~\bibnamefont{Kirkpatrick}},
  \bibinfo{journal}{Advances in Physics} \textbf{\bibinfo{volume}{20}},
  \bibinfo{pages}{325} (\bibinfo{year}{1971}{\natexlab{b}}).

\bibitem[{\citenamefont{Koll\'ar et~al.}(2012)\citenamefont{Koll\'ar, Kiss,
  Novotn\'y, and Jex}}]{bib:PhysRevLett.108.230505}
\bibinfo{author}{\bibfnamefont{B.}~\bibnamefont{Koll\'ar}},
  \bibinfo{author}{\bibfnamefont{T.}~\bibnamefont{Kiss}},
  \bibinfo{author}{\bibfnamefont{J.}~\bibnamefont{Novotn\'y}},
  \bibnamefont{and} \bibinfo{author}{\bibfnamefont{I.}~\bibnamefont{Jex}},
  \bibinfo{journal}{Phys. Rev. Lett.} \textbf{\bibinfo{volume}{108}},
  \bibinfo{pages}{230505} (\bibinfo{year}{2012}).

\bibitem[{\citenamefont{Leung et~al.}(2010)\citenamefont{Leung, Knott, Bailey,
  and Kendon}}]{bib:VivKendon2010}
\bibinfo{author}{\bibfnamefont{G.}~\bibnamefont{Leung}},
  \bibinfo{author}{\bibfnamefont{P.}~\bibnamefont{Knott}},
  \bibinfo{author}{\bibfnamefont{J.}~\bibnamefont{Bailey}}, \bibnamefont{and}
  \bibinfo{author}{\bibfnamefont{V.}~\bibnamefont{Kendon}},
  \bibinfo{journal}{New Journal of Physics} \textbf{\bibinfo{volume}{12}},
  \bibinfo{pages}{123018} (\bibinfo{year}{2010}).

\end{thebibliography}
\end{document}